\newif\ifabstract
\abstracttrue
 \abstractfalse 
\newif\iffull
\ifabstract \fullfalse \else \fulltrue \fi

\documentclass[11pt]{article}
\usepackage{amsfonts}
\usepackage{amssymb}
\usepackage{amstext}
\usepackage{amsmath}
\usepackage{xspace}
\usepackage{theorem}
\usepackage{graphicx}
\usepackage{url}
\usepackage{hyperref}
\hypersetup{
    colorlinks=true,
    linkcolor=blue,
    filecolor=magenta,      
    urlcolor=red,
    pdftitle={Overleaf Example},
    pdfpagemode=FullScreen,
    }

\usepackage{graphics}
\usepackage{colordvi}
\usepackage{colordvi}
\usepackage[dvipsnames]{xcolor}
\usepackage{titling}
\usepackage{subfigure}
\usepackage{authblk}
\usepackage[utf8]{inputenc}
\usepackage[backend=biber,style=numeric,sorting=none]{biblatex}
\advance \oddsidemargin by -0.8in
\newcommand{\myparskip}{3pt}
\parskip \myparskip
\date{}

\begin{document}

\title{Published in \textbf{Scientific Reports}, Vol.\textbf{12}, Issue-1, Article No.\textbf{14674}, Aug 2022.\\
DOI: \href{https://www.nature.com/articles/s41598-022-18701-y.pdf}{10.1038/s41598-022-18701-y}.
\\Ultrafast laser inscribed waveguides in tailored fluoride glasses: An enabling technology for mid-infrared integrated photonics devices}

\author[1,*]{T Toney Fernandez}
\author[1]{B Johnston}
\author[1]{S Gross}
\author[2]{S Cozic}
\author[2]{M Poulain}
\author[3]{H Mahmodi}
\author[3]{I Kabakova}
\author[1]{M Withford}
\author[1]{A Fuerbach}
\affil[1]{MQ Photonics Research Centre, School of Mathematical and Physical Sciences, Macquarie University, NSW, 2109, Australia}
\affil[2] {Le Verre Fluoré, 1 rue Gabriel Voisin - Campus KerLann, F-35170 Bruz, Brittany, France}
\affil[3] {School of Mathematical and Physical Sciences, University of Technology Sydney, Ultimo, NSW, 2007, Australia}
\affil[*]{Corresponding author: toney.teddyfernandez@mq.edu.au}

\begin{titlepage}
\maketitle

\thispagestyle{empty}

\begin{abstract}
Zirconium fluoride (ZBLAN) glass, the standard material used in fiber-based mid-infrared photonics, has been re-designed to enable the fabrication of high index-contrast low-loss waveguides via femtosecond laser direct writing. We demonstrate that in contrast to pure ZBLAN, a positive index change of close to 10$^{-2}$ can be induced in hybrid zirconium/hafnium (Z/HBLAN) glasses during ultrafast laser inscription and show that this can be explained by an electron cloud distortion effect that is driven by the existence of two glass formers with contrasting polarizability. High numerical aperture (NA) type-I waveguides that support a well confined 3.1 $\mu$m wavelength mode with a mode-field diameter (MFD) as small as 12 $\mu$m have successfully been fabricated. These findings open the door for the fabrication of mid-infrared integrated photonic devices that can readily be pigtailed to existing ZBLAN fibers. 
\end{abstract}

\end{titlepage}

\label{--------------------------------------------sec: intro---------------------------------------------------}
\section{Introduction}\label{sec: intro}

Optical fibre technology based on silica glass has revolutionised application fields as diverse as telecommunications and manufacturing (e.g. laser cutting and welding) by providing a robust and efficient integrated platform for the generation of visible and near-infrared light. However, for wavelengths longer than about 2.5 $\mu$m, silica fibres become virtually opaque and alternative soft glass materials must be used. Over the past years, fluoride fibre technology based on ZBLAN glass~\cite{POULAIN1973} has shown great promise and has now finally reached a stage of maturity where it is poised to initiate a similar disruption in the mid-infrared~\cite{Jackson2012}. For example, fibre-based supercontinuum sources in the mid-infrared are capable of generating electromagnetic radiation with a wavelength coverage and the brightness of a synchrotron, yet with the footprint of a table-top instrument~\cite{Hudson17}, thus enabling fast spectral mapping with a signal-to-noise ratio (SNR) that surpasses that achievable with a synchrotron source and in a shorter acquisition time~\cite{Borondics18}. However, in order for mid-infrared technology to become a truly disruptive technology, the development of field-deployable systems, i.e. systems that are able to operate under harsh and sometimes even extreme environmental conditions, in stark contrast to purely laboratory-based proof-of-principle instruments, is required. A prerequisite for this is the availability of connectorized (i.e. fiber pigtailed) and thus compact and robust integrated optical components such as splitters, couplers, circulators, and wavelength selective elements, to only name a few. While all these are readily available “off-the-shelf” for silica-glass based systems operating in the near-IR, equivalent components for the mid-infrared are still largely missing due to challenges of high thermal expansion, hygroscopicity and steep viscosity-temperature curve for most mid infrared materials including fluorides. Also, to date, splicing device manufacturers do not offer equipments fully dedicated to soft glasses hence, it is difficult to obtain the high temperature control required around 250-350$^{\circ}$C to process in optimal conditions with fluoride fibers~\cite{Rowe, Schafer18}. In this work, we present a potential solution for this fundamental problem. 
\\Ultrafast Laser Inscription (ULI) is a well-studied and utilised technique for the fabrication of buried optical waveguides inside various different glasses~\cite{Gattass}. While the method has the potential to solve the “mid-infrared bottleneck”, standard ZBLAN glass has been shown to respond with only a very limited induced positive/negative index change during ULI, and as such virtually all reported ULI ZBLAN devices are based on a depressed cladding inscription approach, resulting in low NA, large mode area guiding~\cite{Gross15,Berube13}.  These structures have been used successfully to produce waveguide lasers in active ZBLAN glasses~\cite{Gross_nanoph}, but are of limited utility in realising other optical components where low loss and mode matching to high NA mid-infrared optical fibres is required. An exhaustive detail of all the techniques used to tailor the refractive index within the ZBLAN glass including slit shaping technique can be found in ref~\cite{Gross2012}.
Other mid-infrared transparent materials like lead-germanates ~\cite{Mamoona2021}, gallo-germanates~\cite{Berube2017}, tellurites~\cite{Smayev2018} and chalcogenides~\cite{Gretzinger2015,Rodenas2012} have been used as substrates for laser inscribed waveguides in this wavelength region. Within this group, gallium lanthanum suphide (GLS) glass is the most attractive glass in demonstrating low-loss waveguiding at longer wavelengths (\textgreater 3 $\mu$m)~\cite{Arriola2017}. But all these glasses have inherently high refractive index which introduces high coupling losses to the lower index fluoride fiber architectures. Even if an intermediate stage is designed to reduce the coupling losses, the large difference in coefficient of thermal expansion (CTE) between the said materials and the fluoride fibers introduces additional thermal management problems for high power applications. The CTE of a ZBLAN glass is $\approx$ 18 $\times$ 10$^{-8}$ K$^{-1}$~\cite{Poulain2010}, whereas for GLS it is more than two orders higher $\approx$ 6 $\times$ 10$^{-6}$ K$^{-1}$~\cite{Hewak2010}.\\
Our goal was thus to develop a mid-infrared compatible glass composition that can produce a smooth, strong and positive refractive index change upon irradiation with ultrafast laser pulses which could then be easily integrated to the existing fluoride fiber architecture. Combined with an optimized inscription strategy, this glass could then be utilized for the inscription of waveguides with mode field diameters and V-numbers (V=$\frac{2\pi}{\lambda}$ a $\times$ NA, where a is the waveguide radius) that perfectly match those of existing optical fluoride fibers, thus enabling the fabrication of fiber-pigtailed integrated components. A recent communication from Heck et.al~\cite{Heck18} reported a positive index change within a fluoride optical fiber upon irradiation with femtosecond laser pulses. Their findings, specifically the significant increase in positive index change when the inscription was carried out at the interface between the core and cladding materials of the fiber was unexplained, yet intriguing as it suggested that ULI in these glasses has the potential to be compositionally tailored. In the case of a ZBLAN optical fiber, a small mol$\%$ of zirconium (Zr) is substituted by hafnium (Hf) in the cladding to lower the refractive index of the glass.  One of the conclusions to be drawn from the findings of Heck et al was thus that the compositional variation between zirconium (Zr) and hafnium in the core and cladding was responsible for the unexpected response to femtosecond laser irradiation, giving rise to a small region of positive index change.  We have reported in the past that strong thermal and compositional concentration gradients (cf section 5.3 in~\cite{Fernandez2018}) could be two triggering factors for refractive index change upon irradiation with femtosecond laser pulses. We speculate that these also played a role in Heck et al’s findings, as aberrated focusing and thermal profiles~\cite{Fernandez_2015} due to inscribing through the curved fiber along with a step-concentration gradient across the core-clad interface, aided in producing the increase in index change reported.
\\In a recent communication, we had empirically predicted, for other families of glass such as the boroalumino silicates, that if the main glass forming element is accompanied by a second glass forming element that is matched in its polarizability field strength (such as the interplay between aluminium and calcium in a silica glass)~\cite{Fernandez2020}, there is a high probability of obtaining waveguides with an improved positive refractive index contrast under ULI~\cite{Fernandez2020}.  In this current work, we have thus explored whether there is a valid analogy in fluoride glasses, by studying ULI in modified ZBLAN compositions containing significant hafnium content. We show that a precise and selective addition of hafnium enables the inscription of optical waveguides with high index-contrast in fluoride glass, while the intrinsically good optical, chemical and mechanical properties are not negatively affected by this compositional re-design.

\section{Results and discussion}\label{sec: R&D}
\begin{figure}[ht]
\centering
\includegraphics[trim={0 0 0 0},width=12.5 cm]{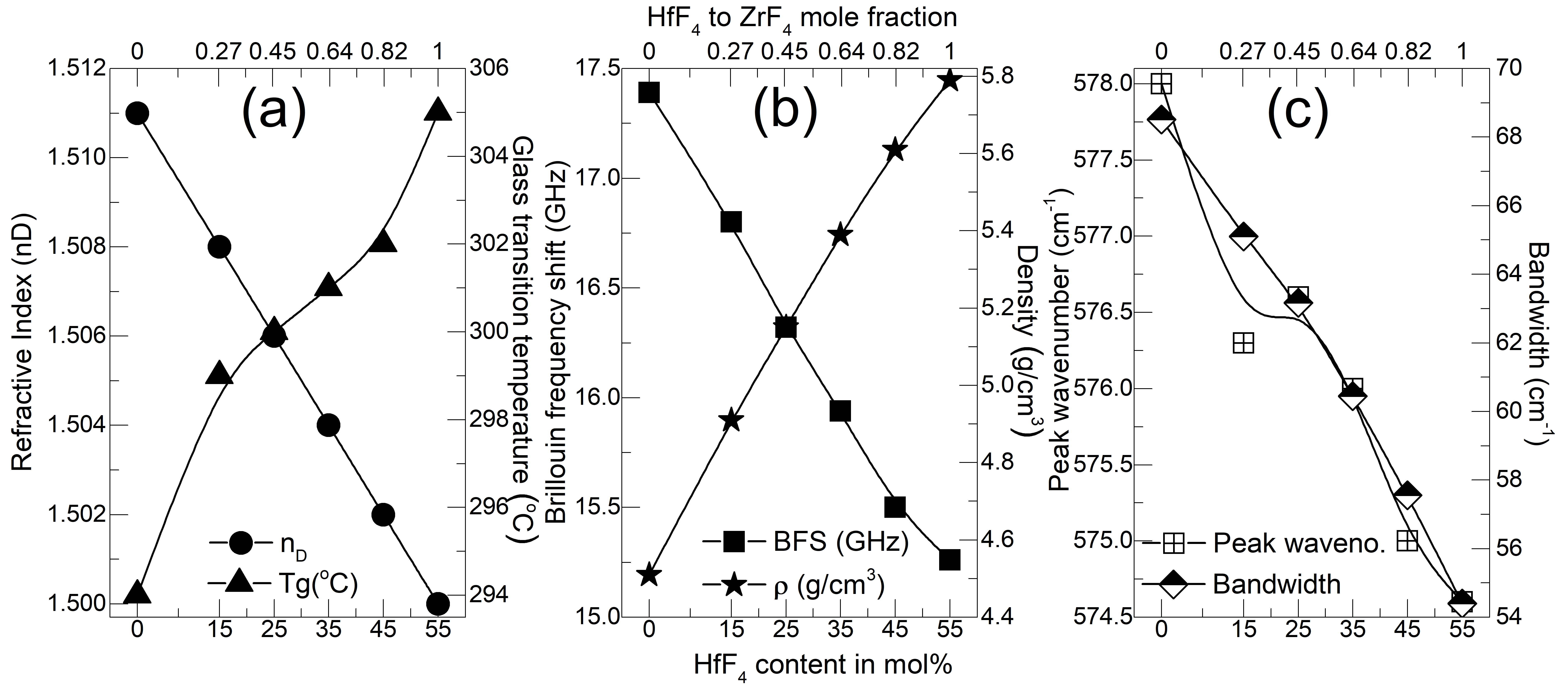}\caption{(a) refractive index (n) (left axis), glass transition temperature (Tg) (right axis) (b) Brillouin frequency shift (BFS) (left axis), density ($\rho$) (right axis) and (c) terminal fluorine Raman peak vibrational frequency (left axis), terminal fluorine Raman peak bandwidth (right axis).}
\label{Fig-1}
\end{figure}

\begin{figure}[ht]
\centering
\includegraphics[trim={0 0 0 0},width=12 cm]{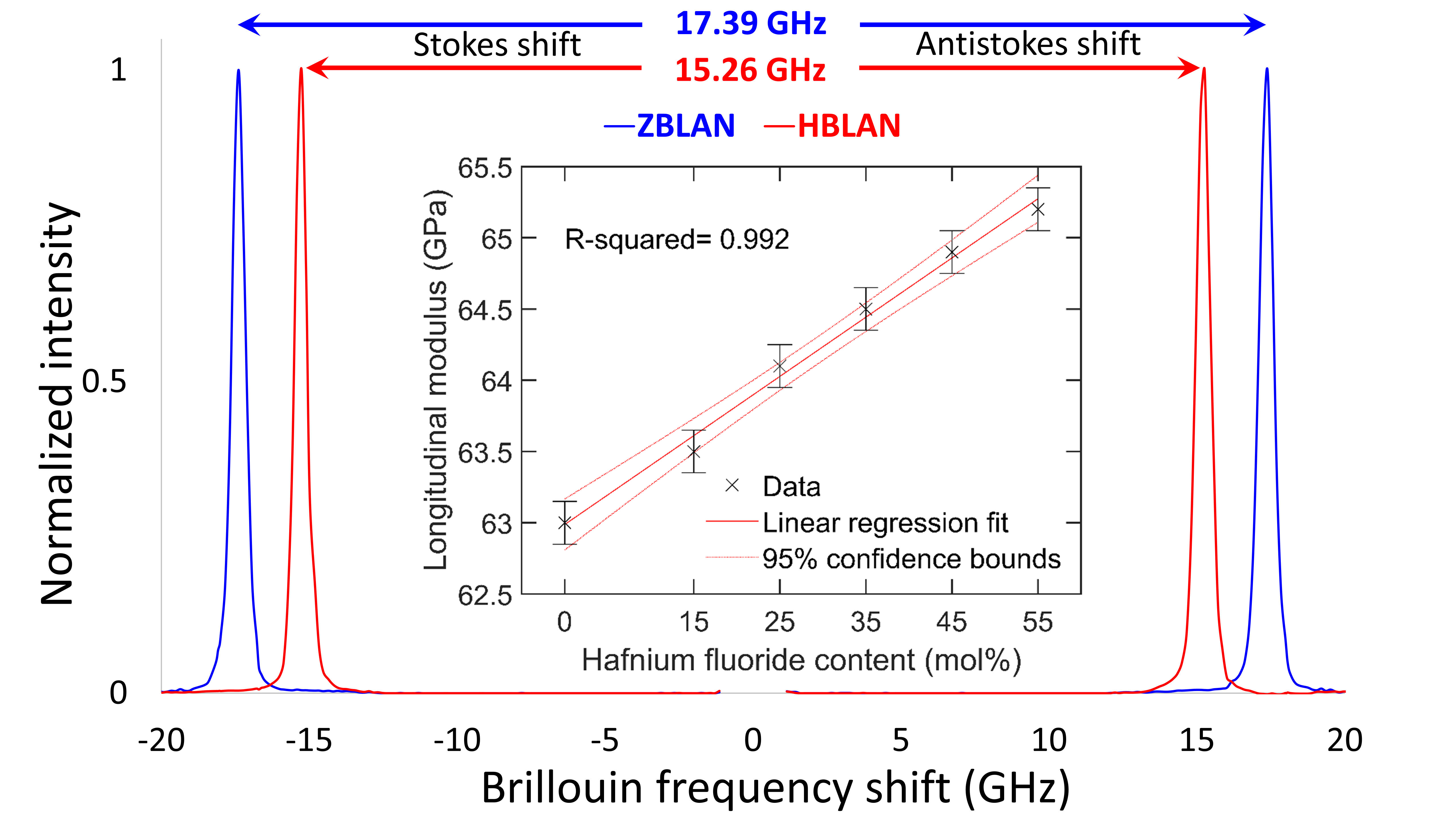}\caption{Brillouin frequency shift spectra for ZBLAN and HBLAN glasses. Inset: The longitudinal modulus for all custom glasses calculated based on Brillouin scattering measurements.}
\label{BFS}
\end{figure}

\begin{figure}[ht]
\vspace{1cm}
\centering
\includegraphics[trim={0 0 0 0},width=10 cm]{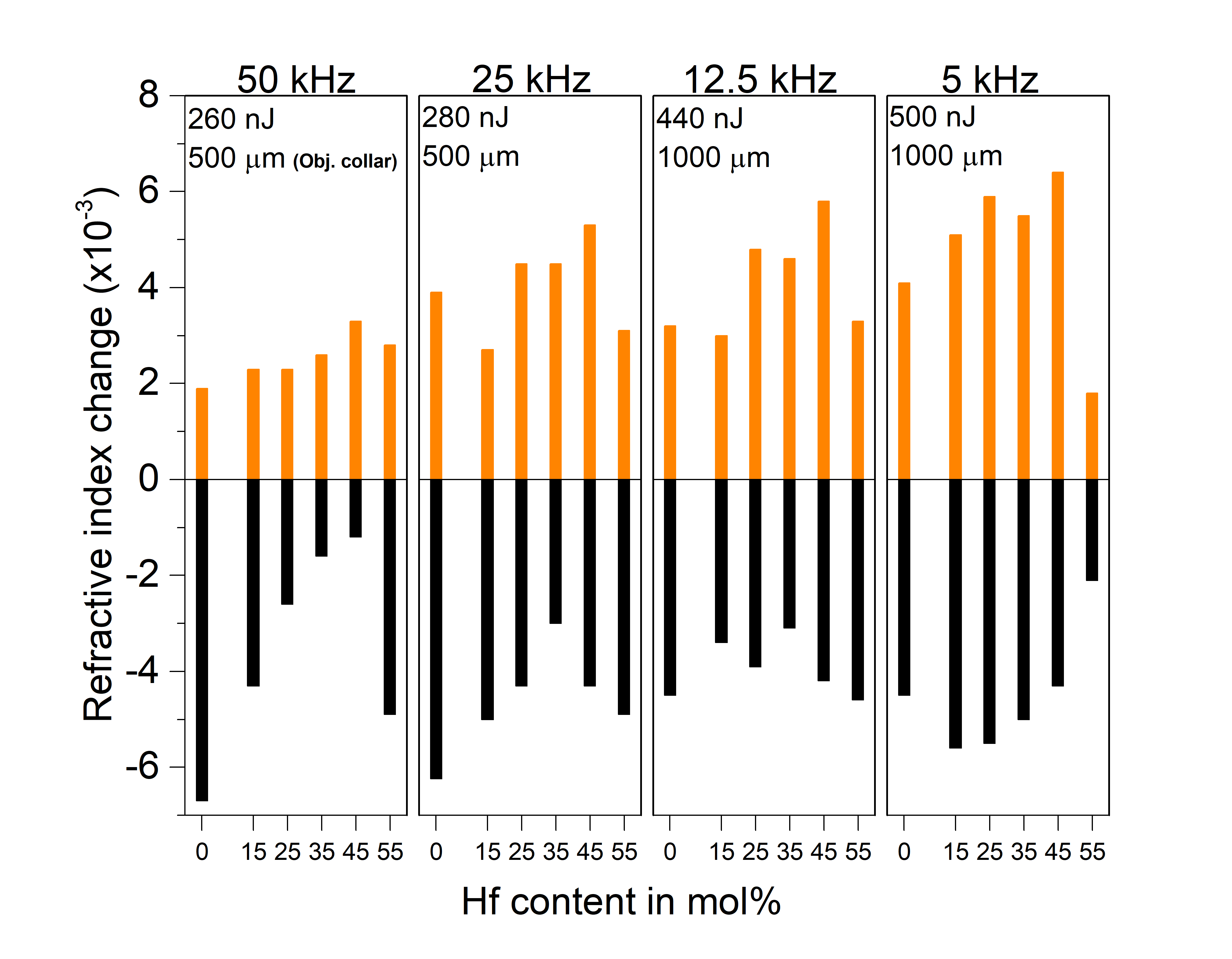}\caption{Evolution of refractive index as a function of HF content for 4 different repetition rates. Laser energy and focussing objective collar was adjusted to achieve maximum index change. Feed rate was kept constant at 0.04mm/s.}
\label{Rep_Rate}
\end{figure}

We fabricated six different glass samples with various hafnium/zirconium content. Further details can be found in the Materials and Methods section. Fig.~\ref{Fig-1} shows that noticeable variations in refractive index (n$_D$), glass transition temperature (T$_g$), Brillouin frequency shift (BFS) and density ($\rho$) were observed in these bulk glasses. It should be noted at this point that there is a large chemical similarity between the hafnium and zirconium atoms and all reported fluorozirconates and fluorohafnates demonstrated strong isomorphism due to same crystalline cell structure and coordination number except for a slightly shorter Hf-F bonds compared to Zr-F bonds. For this reason, hafnium is commonly used for adjusting the refractive indices of the core and cladding glasses in fluoride optical fibers.
\\Raman spectra of all samples and discussions are provided in the supplementary document. The most noticeable variation was found in the distinctive vibrational frequency peak which decreased from 578 $cm^{-1}$ to 574.6 $cm^{-1}$ as the HfF$_4$ content was increased in the composition. As this peak comes from the terminal fluoride vibration due to the heavy stagnant Zr and Hf atoms, a variation in its frequency is believed to be associated with counter cation (Ba, La, Al and Na) re-arrangement~\cite{Phifer91,Gross13}. Understanding the basis of these variations are key in interpreting the origin of index change due to ultrafast laser-matter interaction. 
\\The molar substitution of HfF$_4$ for ZrF$_4$ explains the monotonic increase in density due to the heavier Hf atoms. But the counter intuitive monotonic decrease in refractive index with increasing (heavier) hafnium content is due to the lower atomic polarizability~\cite{Ghosh02} as a result of lanthanide contraction. Hafnium, whose inner 4f orbital is large and diffuse, fail to shield the valence shell from the attraction of the nucleus. Such a strong attractive force on the valence electrons causes a contraction in the size of the electron charge cloud reducing its ability to be distorted when interacted with an electromagnetic wave.  This further explains the higher glass transition temperatures for higher hafnium content glasses since the bonding between two relatively low polarizable atoms like hafnium and fluorine (lowest among halides) give rise to a greater accumulation of electron density in the bonding region. The attractive forces of the nuclei acting on the bonding electrons are thus stronger, hence requiring a higher temperature/energy to break them. This is directly evident from the measured Brillouin frequency shift (BFS) values (Fig.~\ref{BFS}) that are higher for ZBLAN (17.39 GHz) as compared to HBLAN (15.26 GHz). The isotropic nature of all the custom glasses was confirmed by the existence of single Stokes and anti-Stokes peaks shown in Fig.~\ref{BFS} as expected from amorphous nature of the material with no long-distance order or symmetry in the glass matrix. The BFS~\cite{Palombo19} in a material depends on the refractive index (n), longitudinal modulus (M) and the physical density ($\rho$) and is given by the relation  $BFS=\frac{2n}{\lambda} \sqrt \frac{M}{\rho}$. When the measured values of BFS, n and $\rho$ are substituted and $\lambda$ being a constant (660 nm), the longitudinal modulus (Fig.~\ref{BFS} inset) linearly increases when ZrF$_4$ is replaced by HfF$_4$, signifying a more rigid bond for the latter and hence a low polarizability.
\\Utilizing the wide parameter space offered by the ultrafast laser inscription it was found that multiscan waveguides fabricated at lower repetition rates ranging between 5-50 kHz, pulse energies between 200 - 700 nJ and feed rates between 0.02 - 0.5mm/s using a 40$\times$, 0.6 NA focussing objective (Olympus, LUCPlan FL N)  were ideal for producing high index contrast waveguides. It was also found that introducing a precise amount of spherical aberration by detuning of the focussing objective collar position between 300 -1500 $\mu$m helped to fine-tune the V-number of the waveguide. 
\\Fig.~\ref{Rep_Rate} shows the incremental evolution of positive and negative index changes in all glasses upon fs-laser waveguide inscription, keeping the feed rate a constant at 0.04 mm/s while the energies and the objective collar position were adjusted to obtain a maximum index change. A single laser scan pass produced a strand of waveguide which is approximately 1 $\mu$m wide and 12-14  $\mu$m tall. Hence to produce multiscan waveguides, 13 laterally shifted passes (pitch of 0.55 $\mu$m) were carried out at a depth of 170 $\mu$m below the surface to create a $\approx$7 $\mu$m wide waveguide.

\begin{figure}[ht]
\centering
\includegraphics[trim={0 0 0 0},width=12 cm]{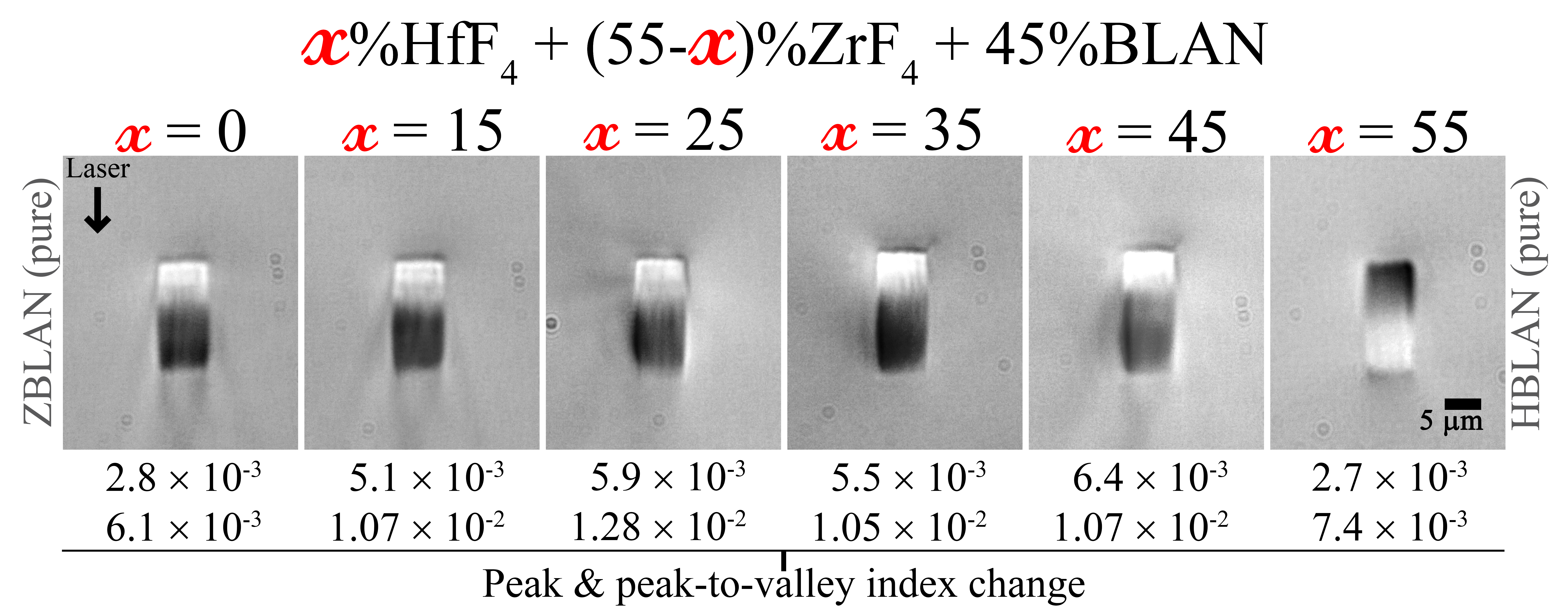}\caption{DIC images of waveguides written in all compositions at 5 kHz, 500 nJ, 0.04 mm/s and focussing objective collar set to 1000 $\mu$m.}
\label{DIC}
\end{figure}

\begin{figure}[ht]
\centering
\includegraphics[trim={0 0 0 0},width=8 cm]{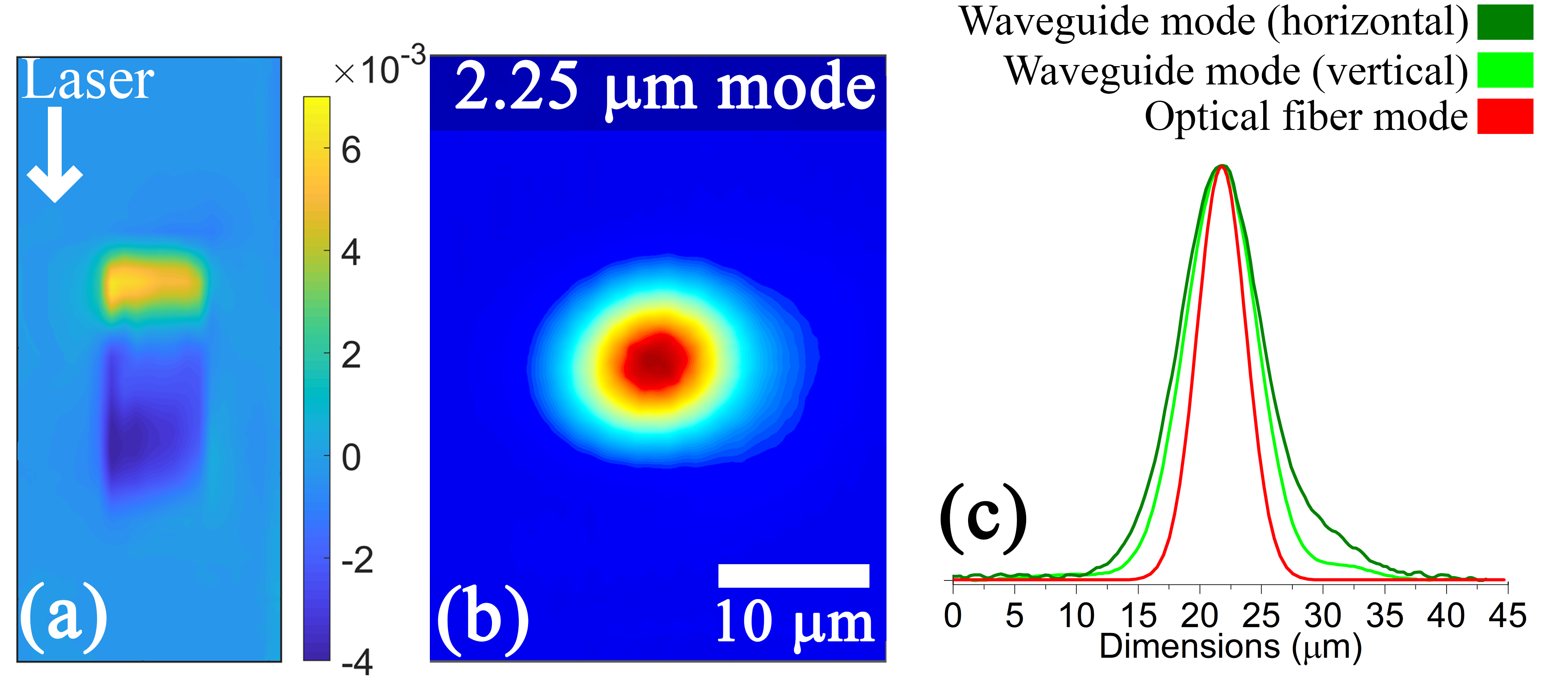}\caption{(a) Refractive index profile (b) 2.25 $\mu$m laser mode and (c) its vertical and horizontal line profiles of the waveguide written in the 45HfF$_4$-10ZrF$_4$-45BLAN glass. DIC image and writing parameters as in Figure 4.}
\label{2.25mode}
\end{figure}

\begin{figure}[ht]
\centering
\includegraphics[trim={0 0 0 0},width=8.5 cm]{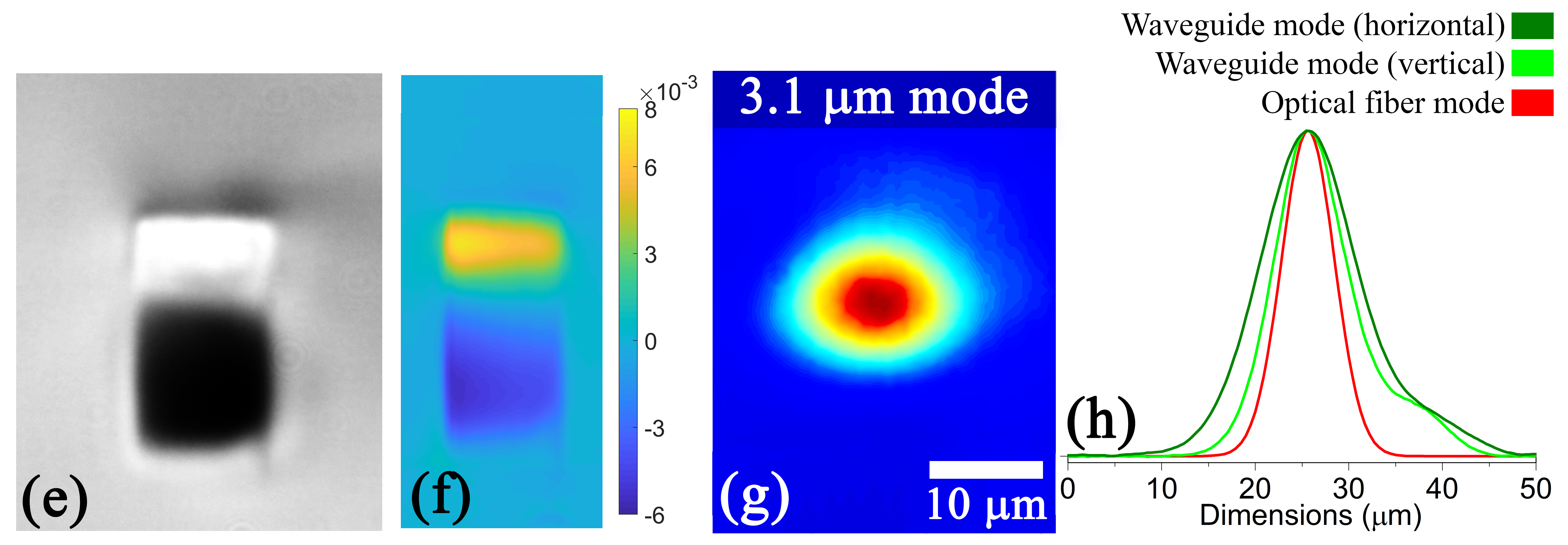}\caption{(a) DIC microscope image (b) Refractive index profile (c) 3.1 $\mu$m mode and (d) its vertical and horizontal line profiles of a 12 $\mu$m wide waveguide (0.6 $\mu$m pitch) written at a repetition rate of 5 kHz and 700 nJ pulse energy in the 45HfF$_4$-10ZrF$_4$ glass. }
\label{3.1mode}
\end{figure}

Fig.~\ref{DIC} shows DIC (Differential Interference Contrast) microscope images of the waveguides written at 5 kHz, 500 nJ, 0.04 mm/s feed rate with focusing objective collar set at 1000 $\mu$m in all six glass compositions. Featured generally with strong positive and a negative index changes, an inversion of index change was also observed in pure HBLAN glass compared to the rest. A 1:3 ratio of positive to negative index change area was found in all waveguides, whereas for HBLAN this was inverted too. This indicates a structural or a compositional change rather than inversion of the laser induced thermal profile~\cite{Fernandez_2015,Song2011}. Hybrid glasses that contains both HfF$_4$ and ZrF$_4$ were found to produce higher positive index changes in comparison to pure HBLAN and ZBLAN glasses.
\\A 2.25 $\mu$m laser mode was guided through the 7 $\mu$m waveguide written in the 45HfF$_4$-10ZrF$_4$-45BLAN (mol$\%$) glass which had the highest positive index change. The guided mode profiles along with 2D refractive index profile are provided in Fig.~\ref{2.25mode}. To compare with the most challenging case, the fiber from Le Verre Fluoré that was used in comparison has a standard NA of 0.23 with a core diameter of 6.5 $\mu$m and a single mode cut off at 1.95 $\mu$m wavelength. The 2.25 $\mu$m waveguide mode dimensions were 11.6 $\times$ 15.3 $\mu$m in comparison to the 10.9 $\mu$m input fiber mode, producing a net coupling loss of 0.26 dB/facet including Fresnel losses (0.18 dB/facet). A further optimization was carried out specifically on the 45HfF$_4$-10ZrF$_4$ glass to provide an additional 20\% increase in refractive index contrast by inscribing at a faster feed rate (0.3 mm/s) with objective collar position set at 1500 $\mu$m. This helped to increase the guiding wavelength beyond 3 $\mu$m. A 12 $\mu$m wide waveguide (0.6 $\mu$m pitch) written at a rep. rate of 5 kHz and 700 nJ pulse energy produced an index contrast of 1.2 × 10-2 which is the highest reported value in a fluoride glass till date. The DIC microscope image of this waveguide along with the 2D refractive index profile and the guided mode profile at 3.1 $\mu$m are provided in Fig.~\ref{3.1mode}. The waveguide mode dimensions were 17.5 x 23.2 $\mu$m in comparison to the 11.5 $\mu$m input fiber mode at 3.1 $\mu$m. Considering the refractive index matching between the fiber and the waveguide this produced a net coupling loss of 1.37 dB/facet. The loss figures are best reported till date considering the smallest waveguide mode in the mid infrared with added advantage of waveguide being a tunable type-I and same material as the fiber for integration. By further optimisation of the focusing geometry to increase the vertical size of the positive refractive index region, we believe that it is possible to obtain a nearly perfectly circular mode which should further reduce the coupling loss.

\begin{figure}[ht]
\centering
\includegraphics[trim={0 0 0 0},width=8 cm]{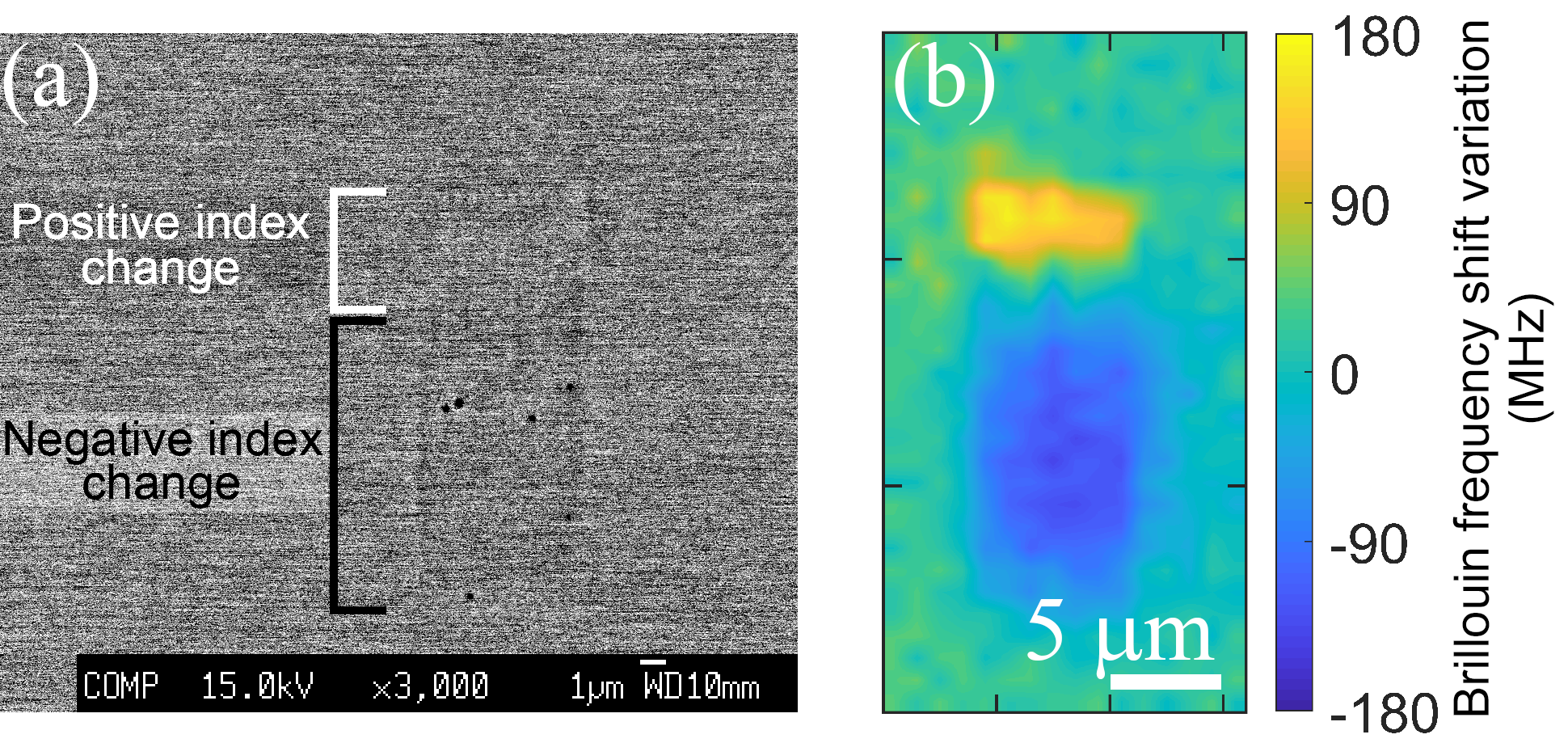}\caption{(a) Backscattered electron microscope image of the 7 $\mu$m wide waveguide written in the 45HfF$_4$-10ZrF$_4$-45BLAN glass. DIC image and writing parameters as in Figure.4. Dark round spots within the negative index change zone are the nano-voids. (b) Brillouin frequency shift mapped across the same waveguide }
\label{EPMA-BFS}
\end{figure}

\begin{figure}[ht]
\centering
\includegraphics[trim={0 0 0 0},width=13 cm]{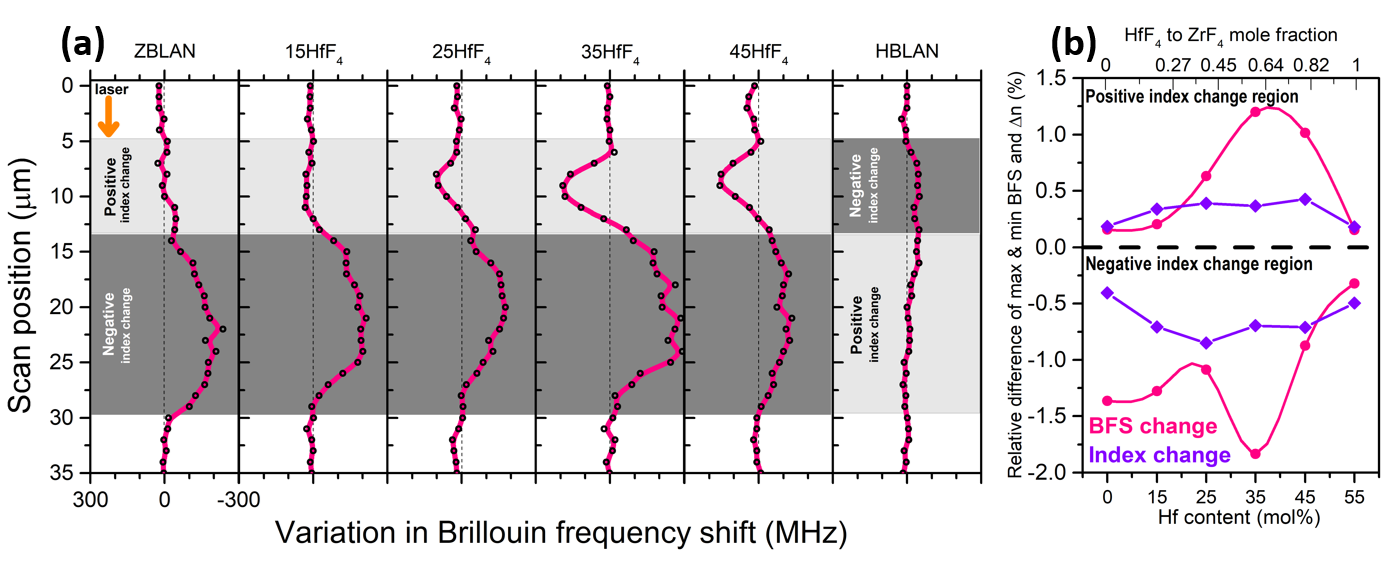}
     \caption{(a) Brillouin frequency shift line scans of waveguides written in all compositions at 5 kHz, 500 nJ, 0.04 mm/s and focusing objective collar set at 1000 $\mu$m. (b) A comparison of relative difference observed between the maximum and minimum values of Brillouin frequency shift and the refractive index change measured in the same waveguides.}
\label{BFS-linescans}
\end{figure}

In order to fine-tune the V-number and thus to further lower the coupling losses, the physical origin of the index change must be identified.  A backscattered electron microscope image of the 7 $\mu$m waveguide written at 0.04 mm/s feed-rate is shown in Fig.~\ref{EPMA-BFS}(a) and shows negligible z-contrast, indicating weak/negligible material densification/rarefaction. The formation of nano-voids (observed as dark round spots) were found exclusively in the negative index change region. Nano-voids/nano-pores are a common feature in the athermal regime of ultrafast laser irradiation and are formed due to free carrier accumulation where there is the highest electron density~\cite{Dai2016}. Their presence exclusively in the negative index change region indicates the existence of a strong gradient profile for laser energy deposition within the glass due to the detuning of the focussing objective collar~\cite{Fernandez_2015, Song2011}. An elemental mapping of all the constituent elements within the glass was featureless, indicating that the formation of waveguides is purely caused by a structural reorganization and fixed stoichiometry. 
Since fluoride glasses are expected to have a higher ionic character~\cite{REAU1989} which 2-4 orders higher than oxide glasses (depending on alkali free or alkali rich composition)  mainly because of the presence of monovalent F- ion, electron beam induced migration~\cite{Jiang2004} should be expected while using any e-beam characterization techniques. Hence, supplementary confirmation was sought through Brillouin scattering measurements across the waveguides as it is a purely light-based probing method of the laser-modified zones. Fig.~\ref{EPMA-BFS}(b) shows the relative Brillouin shift across the laser modified zones with respect to the bulk. Moderate shifts of +172 MHz and \textminus135 MHz were observed in the positive and negative index change regions of the waveguide, respectively. Compared to the measurement uncertainty (10 MHz), estimated based on 10 measurements across three unmodified glasses, this change is statistically significant. It is worth pointing out that the overall difference of the Brillouin frequency shifts in this glass for the whole range of compositions is found to be 4.25 GHz (Figure 2). We could thus confirm the results found from electron probe micro analysis (EPMA) that the waveguide formation is not due to ion migration or any local stoichiometry changes. BFS line scans as shown in Fig.~\ref{BFS-linescans}(a) were carried out from top to bottom along the incoming fs-laser direction in those waveguides whose DIC images are shown in Fig.~\ref{DIC} revealing that the magnitude of Brillouin shift reaches the maximum for glasses modified with 35HfF$_4$-20ZrF$_4$-45BLAN. It is interesting to note that the shift is quite sensitive in the negative index zones whereas the positive index zones are featureless for both pure ZBLAN and HBLAN glasses. We could deduce that the negative index change zone formation is based on a structural modification leading to rarefaction. The formation of nano-voids exclusively at the negative index change zone supports this argument. A comparison of the relative difference observed between the maximum and minimum values of Brillouin frequency shift and the refractive index change measured in the respective waveguides is shown in Fig.~\ref{BFS-linescans}(b). First and last data points of pure HBLAN and ZBLAN glass waveguides indicate that the relative change in BFS and $\Delta$n are the same in the positive index change zone. Since $BFS=\frac{2n}{\lambda} \sqrt \frac{M}{\rho}$ , this substantiates a lack of change in density (supplemented by backscattered electron microscopy (BSE)) and longitudinal modulus in the guiding region during waveguide formation. Whereas both deviates to a maximum value for 35HfF$_4$-20ZrF$_4$-45BLAN composition within the positive index change zone. Given that the density change within the waveguide is negligible, the change in longitudinal modulus must be the responsible factor for this strong deviation. Since the composition contains two glass formers (HfF$_4$ and ZrF$_4$) and the deviations are maximum when the mole fraction (HfF$_4$:ZrF$_4$) is between 0.45-0.82, the responsible factor is believed to be a mixed glass former effect as the rest of the composition (BLAN) is kept constant across all samples. This result is quite surprising because Hf and Zr are highly isomorphic and do not contribute to the glass former effect during the bulk glass formation. 

\begin{figure}[ht]
\centering
\includegraphics[trim={0 0 0 0},width=8 cm]{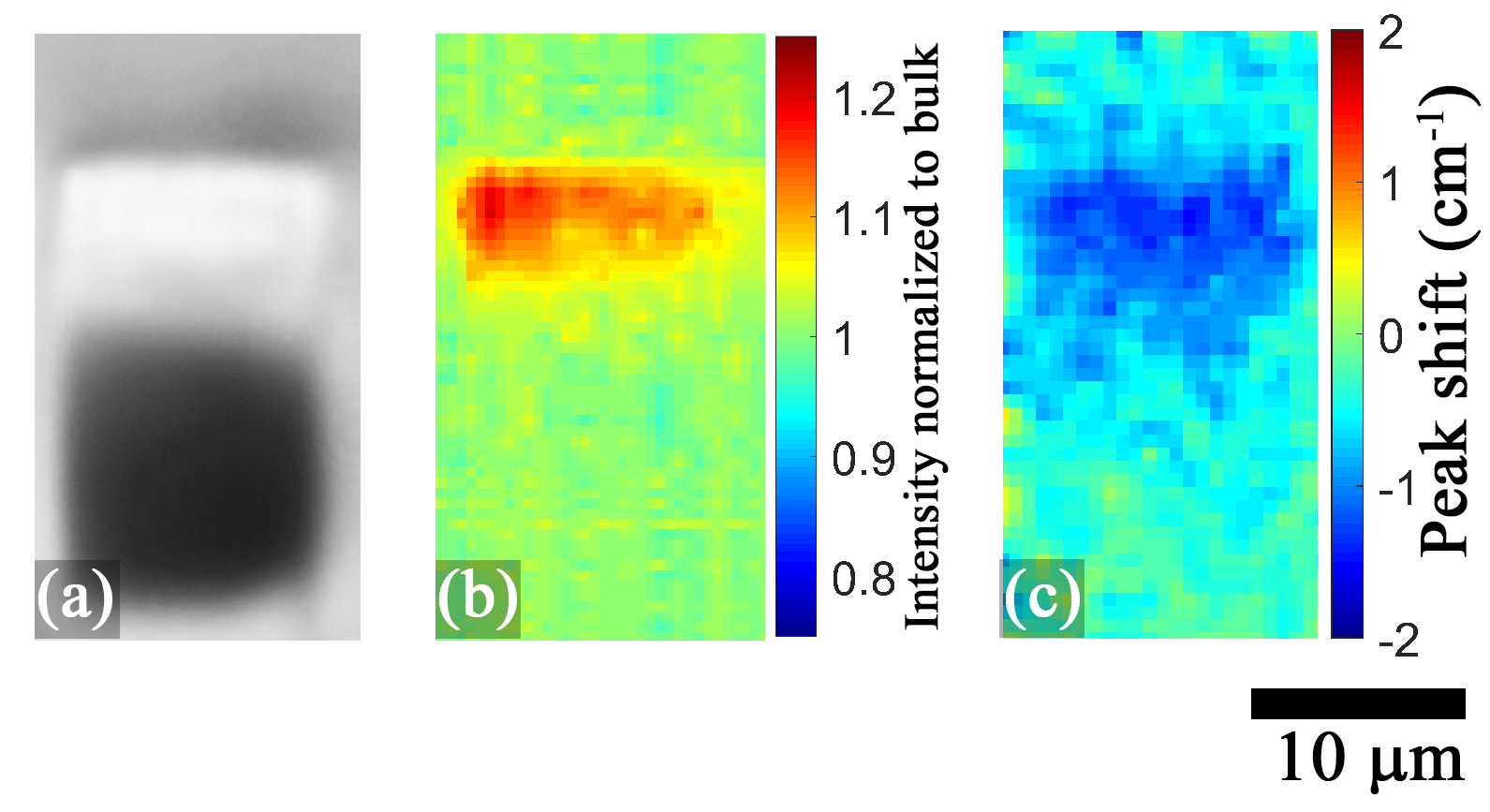}
     \caption{(a) DIC image (b) 575 $cm^{-1}$ Raman peak intensity and (c) its frequency shift of the 12 $\mu$m wide waveguide (0.6 $\mu$m pitch) written at a rep. rate of 5 kHz and 700 nJ pulse energy in the 45HfF$_4$-10ZrF$_4$-45BLAN glass. }
\label{Raman}
\end{figure}

\begin{figure}[ht]
\centering
\includegraphics[trim={0 0 0 0},width=17 cm]{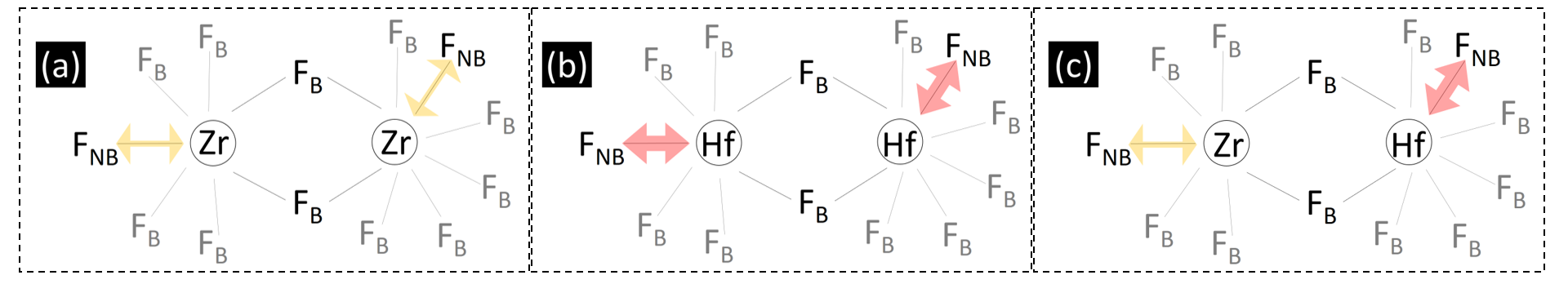}
     \caption{Electron cloud of terminal/non-bridging fluorines(F$_{NB}$) experiencing  (a) moderate distortion from Zr atoms in pure ZBLAN glasses (b) High distortion from Hf atoms in pure HBLAN glasses (c) terminal fluorines with diverse characteristics due to site specificity depending on its immediate neighbour in hybrid glasses.  }
\label{Schematic}
\end{figure}

Typically, when two isomorphic glass formers are mixed within a glass composition, specifically when one mole of glass former is replaced by another at constant glass modifier composition, the effect is non-linear and non-additive which is reflected in a turning point within its optical, thermal and mechanical properties~\cite{Wang2017}. This effect usually peaks when both species have equal content within a composition because during glass formation, occupiable sites for a particular species are not necessarily energetically or morphologically favourable for a second kind of glass former. In other words, if the three dimensional random network could be considered as a potential energy landscape, two different species will see an entirely different 3D energy profile with entirely different potentially occupiable sites~\cite{Dyre2009}. A 50\% relative substitution between the two different species hence shows a largest site mismatch demonstrating the largest contrast to its properties. 
In our case, Hf and Zr being isomorphic can be substituted one for another with linear and additive results as evident from the results of Fig.~\ref{Fig-1}. This possibility makes the HfF$_4$-ZrF$_4$ mixed glass in general difficult to discriminate from each other and identifying the role of each glass former in the overall characteristics of the glass is challenging. However, during waveguide inscription, as it transits through a fast quenching process, the modified zone exhibits a mixed glass former effect, especially for glasses with mole fractions around 0.5. 
Raman mapping revealed no changes of the intensity, peak shift and bandwidth of the majority of spectral peaks. The exception was the main peak attributed to the symmetric stretching vibration of the terminal fluorine bonds attached to Zr or Hf atoms. This was the case for all waveguides written across all compositions and inscription parameters. Fig.~\ref{Raman} is a representative image of the 12 $\mu$m waveguide written at 0.3 mm/s translation speeds in the 45HfF$_4$-10ZrF$_4$ glass. It demonstrates the generation of such terminal fluorine bonds and the reduction of its vibration frequency within the positive index zone. 
Since there is a high disparity in the atomic polarizability between Hf (4.3 Å) and Zr (170.6 Å$^3$) due to the lanthanide contraction effect in the former, the generation of new terminal fluorines is subjected to strong electron cloud distortion depending on the parent atom to which it is attached. The atomic number of Hf (72) is almost the double of Zr (40) but the ionic radius due to lanthanide contraction in Hf (83 Å) is quite similar to Zr (84 Å). This exhibits how much tighter the nucleus of Hf binds its electrons to itself, so the deformation of its electronic shells under an electric field is more difficult. 
 Fluorine atoms also possess a low atomic polarizability due to their smaller size. Hence the electron localization when a terminal fluorine gets attached to Hf is very high compared to Zr. The observation of mixed glass former effect during waveguide formation can be explained due to the formation of terminal fluorines attached to glass formers with highly diverse polarizabilities. After the glass undergoes a fast quenching process initiated by the energy of the fs laser pulses, the formation of terminal fluorines are site specific and congested caused by the physical presence of the surrounding energy-diverse ligands within the three dimensional network. Fig.~\ref{Schematic} represent the very simple case demonstrating the effect on the electron cloud of terminal fluoride attached to (a) Zr-Zr (b) Hf-Hf (c) Zr-Hf molecular units. Red arrow indicates the strong electron cloud distortion due to the low polarizable Hf atom whereas the yellow arrow indicates a relatively small distortion due to the highly polarizable Zr atom.  Hence the glasses with HfF$_4$:ZrF$_4$ $\approx$ 0.5 shows the maximum effect. It further explains the higher refractive index change induced by the distorted electron clouds in glasses that contain both glass formers and the much lower index change in single component pure ZBLAN and HBLAN glasses.

Terminal fluorine bonds do have a higher polarizability compared to the bridged ones, but since the production is the same across all compositions, it is reasonable to believe that the structural rearrangement obeying the mixed glass former effect induces a higher polarizability due to the strong disparity between Hf and Zr.
In summary, high numerical aperture, highly controllable mid-infrared waveguides were fabricated in a redesigned ZBLAN glass. Our results open up the way to the possibility of pigtailing existing fluoride fibers to integrated functional glass chips, thus enabling a new hybrid architecture for the development of fully integrated field-deployable mid-infrared photonic systems. The generation of highly polarizable terminal fluorides due to the fs-laser inscription was identified as the main mechanism for the high positive index change obtained in the guiding region, whereas the low index change region was characterized by structural modifications with nanovoid formation. Laser induced ion migration or stoichiometry change did not contribute to waveguide formation. The maximum refractive index change can be manipulated by varying the ratio of the two glass formers by controlling electron cloud distortion due to fs-laser induced mixed glass former effect. Hence, this work can also serve as a guideline for materials-based optimization of waveguide fabrication in other glasses and for other wavelengths.

\section*{Materials and methods}
Six different glasses with varying HfF$_4$ content starting with a conventional pure ZBLAN glass having a composition of 55 ZrF$_4$ and the rest 45 mol\% comprised of BaF$_2$, LaF$_3$, AlF$_3$, NaF was used. xHfF$_4$-(55-x)ZrF$_4$-45BLAN where x = 0, 15, 25, 35, 45, 55. The stoichiometry of Ba, La, Al and Na is kept constant for all glasses. Glasses were prepared with the conventional melt quenching technique at the Le Verre Fluore industrial facility. To avoid any experimental or alignment errors during waveguide inscription, all samples were mounted on a float glass substrate, polished down to same thickness and flatness before inscription was carried out at the same time on all samples. A Pharos femtosecond laser system operating at a central wavelength of 1030 nm, pulse duration of 240 fs and a variable repetition rate starting from single pulse up to 1 MHz was used for inscription. We found that lower repetition rates provide ideal inscription windows for multiscan waveguides and hence in this work concentrated on values between 5 - 50 kHz. After inscription, 11 mm long waveguides whose end facets were polished were imaged using differential interference contrast microscopy using an Olympus inverted microscope. The refractive index was profiled using Rinck near field refractometer and mode profiled using a Dataray wincam S-WCD-IR-BB-30 beam profiler. Structural and chemical characterization of the waveguides were carried out using Scanning electron microscope on a JEOL JXA-8500F field-emission EPMA and Micro-Raman spectroscopy with 633 nm excitation wavelength on a Renishaw inVia Raman microscope in confocal mode using a 100×objective (spatial resolution $\approx$0.5 $\mu$m). 
Brillouin scattering experiments were carried out to identify differences in the mechanical properties of waveguides and bulk glasses of all compositions. Brillouin spectroscopy is based on inelastic light scattering where photons exchange energy with phonons within the material, leading to a Brillouin frequency shift (BFS, $\Omega$) between the incident and scattered light. This shift is directly proportional to the speed of longitudinal sound waves ($\nu_s$), the refractive index n and inversely proportional to the probing laser wavelength $\lambda$=660 nm, $\Omega$=$\frac{2n}{\lambda}\nu_s$. Thus, Brillouin light scattering directly probes the propagation speed of acoustic phonons together with the material refractive index. The acoustic speed, in turn, is the function of the material mechanical properties, namely its longitudinal elastic modulus M and the material density $\rho$, and given by $\nu_s$=$\sqrt \frac{M}{\rho}$.  Spontaneous Brillouin scattering measurements were carried out as a complementary light based technique to electron microscopy as fluoride glasses have a higher ionic character and can be affected by electron beam induced migration of elements within the characterizing region. Brillouin frequency shifts (BFS) were measured using 660 nm single frequency Cobolt Flamenco laser (HÜBNER Photonics) through a confocal microscope (CM1, TableStable Ltd) and the spectra were collected using a 6-pass tandem scanning Fabry-Perot interferometer (TFP1, TableStable Ltd). The backscattering light was collected by an objective lens (20X Mitutoyo Plan Apo infinity corrected objective, NA=0.42, WD = 20 mm) and redirected to the interferometer for analysis. Both line and 2D mappings were carried out by moving the sample on the 3D microscopy stage (SmarAct) along one and two axes of the stage, while keeping the optical system and the objective lens stationary. This experimental apparatus resulted in measurement with spatial resolution of approximately 2 $\mu$m x 2 $\mu$m x 100 $\mu$m in X-Y-Z direction, respectively. The spectral resolution of our instrument is determined by the distance between the mirrors of Fabry-Perot scanning interferometer (3 mm) and the number of acquisition channels (512) to be approximately 276 MHz. The spectral extinction ratio of Fabry-Perot interferometers is above $10^{10}$~\cite{Sandercock1976}.The acquisition time for each point measurement was 20 s to optimise signal-to-noise-ratio and improve the fitting precision. The raw data collected by the spectrometer is fitted using Damped Harmonic Oscillator (DHO) model for each individual Brillouin peak and the centre of these peaks is what determines the Brillouin frequency shift values reported in this manuscript.

\section*{Data availability}
The authors confirm that the data supporting the findings of this study are available within the article [and/or] its supplementary materials.

\section*{Acknowledgements}
This work is funded by the US Air Force Office of Scientific Research under award number FA2386-19-1-4049. This work was performed in-part at the OptoFab node of the Australian National Fabrication Facility, utilising NCRIS \& NSW state government funding. The authors acknowledge the use of facilities supported by Microscopy Australia at the Electron Microscope Unit within the Mark Wainwright Analytical Centre at UNSW Sydney.

\printbibliography

\end{document}